\documentclass[a4paper, 12pt]{article}
\begin{document}

\title{Obstructions, Extensions and Reductions. \\
Some applications of Cohomology \thanks{To be published in the Proceedings of: 
SYMMETRIES AND GRAVITY IN FIELD THEORY. Workshop in honour of Prof. J. A. de Azcarraga. 
June 9-11, 2003. Salamanca, (Spain)}}
\author{Luis J. Boya \\
 Departamento de F\'{\i}sica Te\'{o}rica, Universidad de Zaragoza. \\ E-50009 Zaragoza, Spain \\
luisjo@unizar.es}

\maketitle

\begin{abstract}
       
		After introducing some cohomology classes as obstructions to orientation and spin
structures etc., we explain some applications of cohomology to physical problems, in especial to
reduced holonomy in $M$- and $F$-theories.

\end{abstract}

\section{Orientation}

	For a topological space $X$, the important objects are the {\em homology groups}, 
$H_{*}(X, A)$, with coefficients $A$ generally in $Z$, the integers. A {\em bundle} 
$\xi: E(M, F)$ is an extension $E$ with fiber $F$ (acted upon by a group $G$) over an space 
$M$, noted $\xi:F \rightarrow E \rightarrow M$ and it is itself a \u{C}ech cohomology element,
 $\xi  \in \hat{H}^{1}(M,G)$. The important objects here the {\em characteristic}
 cohomology {\em classes} $c(\xi) \in H^{*}(M, A)$.  

Let $M$ be a manifold of dimension $n$. Consider a frame $e$ in a patch $U \subset M$,
 i.e. $n$ independent vector fields at any point  in $U$. Two frames $e$, $e'$ in $U$
 define a unique element $g$ of the general linear group $GL(n, R)$ by $e' =g \cdot e$,
 as $GL$ acts freely in $\{ e \}$. An {\em orientation} in $M$ is a global class of frames,
 two frames $e$ (in $U$) and $e'$ (in $U'$) being in the same class if det $g>0$
 where $e'= g \cdot e$ in the overlap of two patches. A manifold is {\em orientable} 
if it is possible to give (globally) an orientation; chosing an orientation 
the manifold becomes {\em oriented}. So the questions are: first, when a manifold
 is orientable and second, if so, how many orientations are there. These questions are 
tailor-made for a cohomological answer. 

This is about the easiest example traslatable in simple cohomological 
terms by obstruction theory. Let $\tau$ be the {\em principal} bundle 
of the tangent bundle to $M$, so the total space $B$ is the set of all 
frames over all points:
\begin{equation} \label{orien}
\tau : GL(n, R) \rightarrow B \rightarrow M
\end{equation}

Matrices in $GL^{+}$ with det $>0$ have index
 two in $GL$, hence are invariant, with $Z_{2}$ as quotient:
 we form therefore an associated bundle $w_{1}=w_{1}(\tau)$:

\begin{equation} \begin{array} {cccccc}
		\ &GL^+& & & & \\
  \ &\downarrow & & & &  \\ 
			\tau : &GL& \rightarrow &B& \rightarrow &M \\
   \     & \downarrow & & \downarrow & & \parallel \\
	   w_1: &Z_2& \rightarrow &B/2& \rightarrow &M   
\end{array} \label{ArrayI}
\end{equation}

Our space $M$ is orientable if the structure group reduces to $GL^{+}$. The set
of principal $G$-bundles over $M$ is noted $\hat{H}^{1}(M,G)$ and
 it is a cohomology {\em set} (in \u{C}ech cohomology) \cite{Cech} . 
Thus $\tau \in \hat {H}^{1}(M, GL)$, and we have associated to $\tau$
 another bundle, name it
 Det $\tau  \equiv  w_{1}(\tau) \in H^{1}(M, Z_{2})$,
called the first Stiefel-Whitney class of $\tau$ (as a real vector bundle). 
The \u{C}ech cohomology set $\hat{H}^1$ becomes a {\em bona fide} abelian group for $G$
abelian, whence we supress the \,$\hat{}$, and the associated bundle, 
still presently principal, becomes a $Z_{2}$ cohomology {\em class}.

Now we have the induced {\em exact cohomology sequence}, i.e.
\begin{equation} \label{3}
 H^0(M, Z_2) \rightarrow \hat{H}^1(M, GL^+)\rightarrow
\hat{H}^1(M, GL)\rightarrow H^1(M, Z_2)
\end{equation}

$\tau$ lives in the third group, and by exactness it has antecedent 
(i.e., $M$ is orientable) if it goes to zero in the final group: the middle bundle above 
reduces if and only if the quotient splits: so we have the result:

\vspace{5 mm}
\hspace{1.5 cm} $M$ is orientable if and only if the first Stiefel-Whitney class of $\tau$, 
that is $w_1 \in H^1(M, Z_2)$, is zero.
\vspace{5 mm}

In other words: $M$ is orientable if $\tau$ reduces its group $GL$ to the connected 
subgroup $GL^+$. According to the fundamental exactness relation, orientability
 means section in the lower bundle, and as it
is principal, the bundle is trivial, hence its class ($w_1$ ) is the zero
 of the cohomology group: the lower bundle in (\ref{ArrayI})  splits.

Alternatively, $M$ is orientable if it has a volume form, which is the same as a
 global frame mod det $> 0$ transformations in overlapping patches. 

	As examples, $RP^2$ is not orientable, but $RP^3$ is orientable, 
where $RP^n$ is the real $n$-dimensional projective space of rays in $R^{n+1}$.
 The reason is the antipodal map $(-1,...,-1)$ in $S^n$ leading to $RP^n = S^n/Z_2$
 is a rotation for $n$ odd, but a reflection for $n$ even; 
note $RP^n$ is not simply connected for any $n$. To have $1$-cohomology
 in any ring the first cohomology group $H^1(M, Z)$ has to be $\neq 0$: 
{\em simply connected spaces are orientable}.

Notice the structure group $GL$ reduces always to the orthogonal group $O = O(n)$:
 any manifold is, in its definition, paracompact, 
and in any paracompact space there are partitions of unity,
 hence for a manifold a Riemann metric is always possible:

\begin{equation} \begin{array} {ccccc}
		O(n)& & & & \\
  \downarrow & & & &  \\ 
	 GL(n, R)& \rightarrow &B& \rightarrow &M \\
         \downarrow & &  & &  \\
	   R^{n(n+1)/2}& \rightarrow &E& \rightarrow &M   
\end{array} \label{ArrayII}
\end{equation}
\vspace{ 1.5 mm}

As the lower row fibre is contractible, the horizontal middle bundle 
lifts; $O(n) \rightarrow  B_0 \rightarrow M$, where $B_0$ is the set of 
orthogonal frames:
 any manifold is riemanizable. Note $O(n)$ is the maximal compact 
subgroup of $GL(n)$, and this is why the quotient is contractible. 
By contrast, not every manifold admits a Lorentzian metric:
it needs to have a Òfield of time-like vectorsÓ globally defined.

Hence the characteristic class $w_1 \in H^1(M, Z_2)$ is the {\em obstruction} to 
orientability:
 it measures if an orientation is possible in a manifold. The next question is: 
If the obstruction is zero, how many orientations are there? Again, the answer
 is written by (\ref{3}): the elements in $\hat{H}^1(M, GL^+)$ falling into $\tau$ are the coset
 labelled by $H^0(M, Z_2)$. In particular, if the manifold is {\em connected},
 $H^0(M, A)=A$,
 and then the zeroth Betti number is $b_0 =1$: hence,  as then  $H^0(M, Z_2)=2 $,

\vspace{ 5mm}
\hspace{ 1.5 cm} A {\em connected orientable} manifold has {\em exactly} two orientations. 
\vspace{ 5 mm}

\section{Spin structure}

 The orthogonal group $O(n)$ is neither connected nor simply connected; 
$0$-connectivity questions lead to the first Stiefel-Whitney class, $1$-connectivity 
to the second (sw) class. Suppose the manifold $M$ is orientable already and write 
the covering group  $Spin(n) \rightarrow SO(n)$:

\begin{equation}
		Z_2   \rightarrow  Spin(n)  \rightarrow  SO(n).                              
\end{equation}

	For $n > 2$ this bundle is the universal covering bundle, as 
  $\pi_1 (SO(n))=Z_2$, $n > 2$. For $n=2$ the spin bundle still covers twice, but 
 now  $\pi_1 (SO(2))=Z$. 

	Endow now $M$ with a riemannian structure (there is {\em no} restriction on doing this,
 see above), and write

\begin{equation} \begin{array} {cccccc}
		\ &Z_2&= &Z_2 & & \\
  \ &\downarrow & &\downarrow & &  \\ 
			\ &Spin(n)& \rightarrow &\tilde{B}& \rightarrow &M \\
   \     & \downarrow & & \downarrow & & \parallel \\
	   \tau : &SO(n)& \rightarrow &B& \rightarrow &M   
\end{array} \label{ArrayIII}
\end{equation}
\vspace{ 3mm}

We say that {\em a manifold admits an spin structure (it is \underline{spinable}) 
if the (rotation) tangent bundle \underline{lifts} to a spin bundle}. 
Again, there is a precise homological answer. From the exact sequence, and with 
 $\tau$ living in third group
\begin{equation} \label{spin}
 H^1 (M, Z_2) \rightarrow \hat{H}^1 (M, Spin(n)) \rightarrow \hat{H}^1(M, SO(n))
\rightarrow H^2(M, Z_2),
\end{equation} 
we see, as before, that if we call $w_2$  the image of $\tau$, there is an 
obstruction to spinability, called the {\em second Stiefel-Whitney class},
\begin{equation}
	                  w_2 = w_2(\tau) \in H^2(M, Z_2)			               
\end{equation}
and a {\em manifold is spinable if and only if} $w_2(M)=0$.
 For example, spheres and genus-$g$ surfaces are spinable.
If the obstruction is zero, how many spin structures there are? Again, a simple 
look at (\ref{spin}) gives the answer: there are as many as $H^1 (M, Z_2)$, which is a 
{\em finite}
 set, of course. For example, for an oriented surface of genus $g$,
 $H^1 (\Sigma_g, Z_2) = Z_{2}^{2g}$, hence $ \#= 2^{2g}$, as is well known in 
string theory; recall that the first Betti number  $b_1(\Sigma_g)= 2g$.

	As examples of spin manifolds, $CP^{2n+1}$ is spinable, but $CP^{2n}$ is not. 
For example, $CP^1 = S^2$, no sw classes; as for $CP^2$, we have that $b_2=1$;
recall also Euler number $(CP^n) = n+1$.

	There is an alternative characterization of spin structures due to Milnor \cite{Milnor},
which avoids using \u{C}ech cohomology sets. From (\ref{ArrayIII})  we have the exact sequence 
(taking values in $Z_2$):
\begin{equation}
     0\rightarrow H^{1}(M, Z_2) \rightarrow H^{1}(B, Z_2) \rightarrow
 H^{1}(SO(n), Z_2) \rightarrow H^2(M, Z_2),         
\end{equation}
and accepting $w_2=0$ we see again the number of spin structures to be 
 $\# H^{1}(M, Z_2)$, as $\tilde {B}$ is the total space of the lifted bundle, 
and lives in the second group.

\section {The first Chern class}

 The Det map $O(n) \rightarrow O(n)/SO(n)=Z_2$ can be performed also
 in complex bundles, with structure group $U$ instead of $O$:

When does a complex bundle $\eta$ reduce to the unimodular group? Write

\begin{equation} \begin{array} {cccccc}
		\ &SU(n)& & & & \\
  \ &\downarrow & & & &  \\ 
			\eta : &U(n)& \rightarrow &B& \rightarrow &M \\
   \     & \downarrow & & \downarrow & & \parallel \\
	   c_1: &U(1)& \rightarrow &B'& \rightarrow &M \\   
\end{array} \label{ArrayIV}
\end{equation}                                                         

	The associated bundle Det $\eta  \in H^{^1}(M, U(1))$ determines the first 
Chern class of $\eta$ by the resolution
  $ Z \rightarrow R \rightarrow U(1)=S^{1}$, as
\begin{equation}
			c_1(\eta) \in H^{2}(M, Z) = H^{1}(M, U(1)).                        
\end{equation}
	$c_1(\eta) =0$ is the condition for reduction to the $SU$ subgroup, 
an important restriction in compactifying spaces in $M$-theory, see later.

	For the general definition of Stiefel-Whitney (and Pontriagin) classes
 of {\em real} vector bundles, and for the Chern classes of { \em complex} 
vector bundles, the insuperable source is \cite{MilnorI}

\section{Euler class as Obstruction}

 A more sophisticated example is provided by the Euler class.
 Look for manifolds $M$ with a global $1$-frame, i.e. a global zeroless vector field: 
let $M$ be orientable; from the coset $S^{n-1}=SO(n)/SO(n-1)$

\begin{equation} \begin{array} {cccccc}
		\ &SO(n-1)& & & & \\
	\ &\downarrow & & & & \\
  \tau: &SO(n) &\rightarrow &B &\rightarrow &M  \\ 
   \     & \downarrow & & \downarrow & & \parallel \\
	  \tau': &S^{n-1}& \rightarrow &B''& \rightarrow &M   
\end{array} \label{ArrayV}
\end{equation} 
                      
	A $1$-frame exists if the (unit) sphere bundle has a section. The last bundle
 produces a map
\begin{equation}
		H^{n-1}(S^{n-1}, R ) \rightarrow  H^{n} (M, R)
\end{equation}

The image of the fundamental class of $H^{n-1}(S^{n-1}, Z)$ is the 
{\em Euler class} $e$ of $M$. The condition of reduction is clearly that $e =0$.
 Here $e[M] = \chi $, the Euler number. The result is the well-known condition for a manifold
 to admit a global $1$-frame: zero Euler number. It is funny (and easy to understand) that
 the theorem has a positive side: you can compute the Euler number by ÒcountingÓ 
the{\em Windungzahl} of the zeros of any vector field, the Poincare-Hopf theorem.

\section{Nonabelian Group Extensions}

 (Cfr. (\cite{AzIz}), Ch. 7). Consider the {\em group extension} problem:
 given groups $K$ and $Q$, find $G$ such that $K \subset G$ normal and $G/K=Q$.  

Recall first the relations, for $Z_H$: center of $H$
\begin{equation}
	H/Z_H \equiv \mbox{Int}\ H  \hspace{1 cm} \mbox{and} \hspace{1 cm}   \mbox{Aut}\ H / \mbox{Int}\ H \equiv 
 \mbox{Out}\ H               
\end{equation}
for  {\em any} group $H$. If there is a solution $G/K=Q$ to our problem, write

\begin{equation} \begin{array} {ccccc}
		\ Z_K& & & & \\
  \downarrow & & & &  \\ 
   K    & \rightarrow &G & \rightarrow &Q   \\
	\downarrow & & \downarrow & &\downarrow  \\
	  \mbox{Int} K  &\rightarrow&\mbox{Aut} K& \rightarrow &\mbox{Out} K \\  
\end{array} \label{ArrayVI}
\end{equation}                                                           
where you {\em construct} the last two vertical arrows. So any extension determines
 a morphism $\alpha : Q \rightarrow  \mbox{Out} K$. Inverse question is: given 
$\alpha \in \mbox{Hom}(Q, Ê\mbox{Out} K)$, are there extensions? How many? Note first, given 
$\alpha$ there is an extension

\begin{equation} \begin{array} {cccccc}
		\alpha: &\mbox{Int}\ K&\rightarrow &X &\rightarrow &\ Q  \\
   &\parallel & &\downarrow & &\alpha \downarrow   \\ 
			&\mbox{Int}\ K   & \rightarrow & \mbox{Aut}\ K& \rightarrow &\mbox{Out}\ K \\       	     
\end{array} \label{ArrayVII}
\end{equation}

	That is

\begin{equation} \begin{array} {ccccc}
		 Z_K & & & & \\
  \downarrow & & & &  \\
		K & & & & \\
\downarrow & & & & \\
	\mbox{Int}\ K & \rightarrow &X& \rightarrow &Q \\  
\end{array} \label{ArrayVII}
\end{equation}

We need to lift the horizontal sequence to have extensions, and we see that there
 is an obstruction in the exact cohomology sequence,

\begin{equation} 
  H^{*}(Q, Z_K) \rightarrow H^{*}(Q, K) \rightarrow H^{*}(Q, \mbox{Int} K) \rightarrow
 H^{*+1}(Q, Z_K)   
\end{equation}

So $\alpha$ produces extensions iff the image of $\alpha$ in the last group is zero:
 {\em this is the obstruction}. Now if we add our knowledge that the abelian extensions
 are given by some $H^{2}$, the obstructions lies in $H^{3}(Q, Z_K)$, and if it is 
zero, the number of extensions is  $H^{2}(Q, Z_K)$; and the obstruction lies in the third group:
\begin{equation}
				w(\alpha) \in H^{3}(Q, Z_K);
\end{equation}
 all this is very similar to the spin or orientation problems.

\section{Structure of Lie groups}

	\indent An unexpected problem where an obstruction {\em is} necessary appears in the existence of simple
 Lie groups. Consider the next simplest group, $SU(3)$. In the natural ${\underline 3}$ representation, 
the group leaves the unit sphere invariant, with $SU(2)$ as little group:
\begin{equation}
	SU(2) = S^3 \rightarrow SU(3) \rightarrow S^5 \subset{R^6} = C^3
\end{equation}
and regard this as a bundle extension; bundles over $n$-spheres are classified by $\pi_{n-1}(G)$, where 
$G$ is the structure group. So here, as
\begin{equation}
\pi_4(S^3)=Z_2
\end{equation}
we have just two solutions, the direct product (which {\em cannot} be the group $SU(3)$, because $S^5$
 is not paralellizable), and the other, necessarily $SU(3)$: {\em the existence of non-trivial budles,
 here (in this case) for nontrivial homotopy classes are crucial for the existence of Lie groups}.
 Incidentally, the map $S^4 \rightarrow S^3$ generating $SU(3)$ is easy to describe: it is the 
{\em suspension of the second Hopf  bundle}. (I thank D. Freed for this remark):

\begin{equation} \begin{array} {crcr}
		\beta : S^1 \rightarrow &S^3 &\rightarrow &S^2  \\ 
			   \      &\ {\scriptstyle \Sigma}  \downarrow &  &\ {\scriptstyle \Sigma} \downarrow  \\
      \ &S^4& \rightarrow & S^3  \\ 
\end{array} \label{ArrayVIII}
\end{equation}

For all Lie groups besides the ``atom in the category", $SU(2)=S^3$ the same obstructions obtain; we leave
the details. It would be nice to invert the question: to {\em deduce} the simple Lie groups from nontrivial
extensions... Incidentally, the Hopf $\beta$ bundle is the {\em second} on the series of higher
 homotopy groups of spheres $\pi_{4n-1}(S^{2n})=Z + ...$, related to the Hopf invariant and to the 
{\em nonexistence} of division algebras besides R, C, H and O \cite{9Adams}. For expressions of 
simple compact Lie groups as {\em finite} twisted products of {\em odd } spheres, see \cite{Boya1}

\section{Special Holonomy manifolds}
Since the advent of $M$-Theory (1995; Townsend, Witten, Polchinski \cite {Mth})
 the problem of compactification of extra dimensions, from 10 to 4 in one extreme to 12 down to 2 
in the other, is becoming more and more acute. If one is a true believer in $M$ (or $F$) theory 
(as I tend to be), this problem is perhaps the central one in physics. The arguments for extra 
dimensions are overwhelming, and so are the reasons why we live in four large dimensions. In a nutshell,
geometric description of nongravitational forces requires extra dimensions, while interactions
transmitted via massless particles do not make physical sense outside four (i.e. the $1/r$ potential 
law).

	Here we want to show, via simple examples, that compactification with extra conditions (like
preserving $N=1$ Supersymmetry) can be easily stated in cohomological terms, as reductions of the
structure/holonomy group of different bundles.

	Consider the ``old" `problem of compactifying the Heterotic String living in $10D$ down to $4D$
. The tangent bundle of the compactifying manifold $K_6$ is
\begin{equation}
\tau : O(6) \rightarrow B \rightarrow M = K_6
\end{equation}

	Now we want $K_6$ to be a manifold orientable ( to integrate), spin (to describe fermions) and with a
(covariant) constant spinor field (to preserve $N=1$ Susy in order to ``understand" the scale of
the Higgs mass together with the existence of chiral fermions). In terms of reduction:

$O(6)$ {\em  reduces}  to  $SO(6)$; $SO(6)$ {\em lifts} to $Spin(6)=SU(4)$. $SU(4)$ {\em reduces} to 
$SU(3)$, which lies in $U(3)$:

	\begin{equation} \begin{array} {cccccc}
		\ &SU(3)& & & & \\
  \ &\downarrow & & & &  \\ 
			\tilde{\tau} : &U(3)& \rightarrow &B& \rightarrow &M \\
   \     & {\scriptstyle det}\downarrow \quad \,  & & \downarrow & & \parallel \\
	   &U(1)& \rightarrow &B'& \rightarrow &M   
\end{array} \label{ArrayVIII}
\end{equation}

	Now $R/Z=S^1$ induces (see above) det $(\hat{\tau} = c_1(\tau (M))$, the first Chern class; hence
{\em M is a complex manifold with $SU(3)$ holonomy, with the first Chern class $=0$, and it can be seen
that this implies the trace of the curvature zero; it is a \underline {Ricci-flat} riemannian 
manifold}: Calabi-Yau manifolds. The search for those manifolds was a prolific industry led by
Phil Candelas in Austin in 1985-92 \cite{Phil}.

	Manifolds with tangent structure groups less than maximal are therefore crucial for $M$-theory;
let us see more examples.

\section{Compactification in $M$-theory}

	Please notice, first, that the {\em reduced holonomy problem} is not the same as 
{\em reducing the structure group}, but in practice both are present together; the link is of course
the two {\em holonomy theorems}: \cite{Koba}

  (1) The structure group can be reduced to the holonomy group, the Ambrose-Singer theorem

 (2) The Lie algebra of the holonomy group is generated by the curvature of the connection
producing the holonomy in the first place, the curvature theorem.

	For a generic riemannian manifold, the possibility of isometry groups and reduced holonomy groups are antagonic:
a generic manifold $M$ has no isometries, and maximal holonomy (e.g. $SO($dim\ $M$) if $M$ orientable).
{\em Viceversa}, special holonomy manifolds have {\em no} isometries in general (what poses a problem for
the existence of gauge groups down in $4D$ by the KK mechanism, see later), and a very symmetric space,
in fact maximally symmetric, like even-dim spheres, has irreducible holonomy $SO(n)$.

	M. Berger (1955) clasified holonomy groups, and came up with several series (like $O(2n) \supset U(n),
O(4n) \supset Sp(n)$ etc.), and just {\em two} (in fact, three; one, corresponding to 
$ Spin(9) \subset SO(16)$ was already known as a symmetric space) special cases:
\begin{equation}
Spin(7) \subset SO(8) \qquad \mbox{and} \qquad G_2 \subset SO(7)
\end{equation}
	
	Both turned out essential in $M$ and $F$ theories. Both come, of course, from the beautiful irreducible
representations provided by the Clifford algebra. Notice this is {\em irreducible} holonomy, in the sense
that the {\em irrep} of the subgroup has the same dim as that of the group, namely $8$ and $7$ bzw.

	To demystify those cases it is enough to ask for those representations of the spin groups $Spin(n)$
which act {\em trans} in the unit sphere; and the answer is, besides the low dimensional cases in which
there are repetitions (like $Spin(6) = SU(4)$), only two more: the \underline {$16$} irrep of $Spin(9)$
and the \underline{$8$} of $Spin(7)$. The first case gives only the Moufang plane $OP^2=F_4/Spin(9)$.
The other is very interesting:

The $8$ irrep of $Spin(7)$ allows for the embeding $Spin(7) \subset SO(8)$. Now $SO(8)$ preserves a
quadratic form. What {\em else} does $Spin(7)$ mantain? What manifolds have $Spin(7)$ holonomy? 
The construction of these manifolds starting by D. Joyce around 1995 \cite{Joyce} has been a great 
achievement. We shall comment on these constructions later, but let us finish first with the $G_2$ case. 

	In general, in a (real or complex) vector space $V$, the set $X$ of $2$-forms and/or quadratic forms
are ``open" in the sense of the orbit space $X/GL(V)$ (\cite {Hitchin}). But $p$-forms dimension grows,
 of course, like $n!/p!(n-p)! > n^2 = \mbox{dim} \  GL(V)$. The exceptions occur naturally in low dimensions:
		A three-form in $6, 7$ and $8$ dimensions and self-dual $4$-form in $8$ dimensions.

	In particular, the group leaving a regular $3$-form invariant in $7$-dim space is $G_2$. 
(Note how the dimensions match: $49 = 7 \times 7 = 35$ (dim $3$-forms) $+14$ (dim $G_2$)). Which 3-form?
 Bilinear forms produce scalar products, but trilinear forms produce an internal law 
$X \times X \rightarrow X$, so one
 guesses that in $R^8$ there is a product! Of course, there is: octonion multiplication. 
The apearance of octonions just reinforces the idea that $M$-theory, as a {\em unique} theory,
has to include octonions, which are {\em unique} structures in mathematics, and responsible for most
of their exceptional objects \cite{Stilwell}. To repeat: the reason of the appearance of octonions in
$M$-theory is this: in the $11$ to $4$ version, the compact manifold has to have $G_2$ holonomy becuse of
supersymmetry;this group preserves a $3$-form, which corresponds to the fully antisymmetric (alternative)
 octonion multiplication $3$-form. No wonder, $G_2$ is the automorphim group of octonions.

In the $F$-theory version, the $12 \rightarrow 4$ descent implies $Spin(7)$ holonomy;
but this group can be seen as unit-octonion ``group" $S^7$ stabilized by $G_2$. In both cases of
 $11=(1, 10)$ dimensions and  $12=(2, 10)$ it is remarkable that supersymmetry unveils octonionion
 structures!

	\section{Structure Diagrams}

It is time to express in diagrams what we are saying. First, there is the structure diagram for $G_2$: 
define it as the little ($=$isotropy) group for the {\em trans} action of $Spin(7)$ in the $7$-sphere 
\cite{9Adams} :

\begin{equation} \begin{array} {ccccc}
	SU(3)&\rightarrow &SU(4) =Spin(6) &\rightarrow &S^7 \\
 \downarrow &   & \downarrow & &   \parallel \\
	 G_2 &\rightarrow &  Spin(7) &\rightarrow & S^7 \\
\downarrow &   & \downarrow & &    \\
S^6 & = &   S^6 & &    
\end{array} \label{ArrayIX}
\end{equation}

So $G_2$ operates in the $6$-sphere of unit imaginary octonions; but $G_2$ is also a subgroup of $SO(7)$, 
witness the {\underline 7} {\em irrep}: the {\em torsion} diagram explains this:

\begin{equation} \begin{array} {ccccc}
		\ & & Z_2 &= & Z_2 \\
  \ & &\downarrow &  &\downarrow   \\ 
		G_2 & \rightarrow &Spin(7)& \rightarrow &S^7 \\
   \parallel     & & \downarrow & & \downarrow \\
 G_2 & \rightarrow & SO(7) &\rightarrow & RP^7  
\end{array} \label{ArrayX}
\end{equation}
and the mixed diagram for $G_2$
\begin{equation} \begin{array} {ccccc}
		SU(2)&= & SU(2) & &\\
  \downarrow & &\downarrow & &  \\ 
		SU(3) &  \rightarrow &G_2& \rightarrow &S^6 \\
    \downarrow & & \downarrow & & \parallel \\
	  S^5 &\rightarrow & V_{11} & \rightarrow &S^6   
\end{array} \label{ArrayXI}
\end{equation}
where $V_{11}$ is a Stiefel manifold, generating the $2$-torsion in $G_2$, in the odd sphere structure
 \cite{Boya2}

\begin{equation}
G_2 = S^3 (\times S^{11}
\end{equation}

Finally, we exhibit the richness of the $S^7$ sphere of unit octonions in the following special holonomy
diagram

\begin{equation}
\begin{array} {ccccccc}
Sp(1)&\subset &SU(3) &\subset &G_2 & \subset & SO(7) \\
\downarrow & & \downarrow & & \downarrow & & \downarrow \\
Spin(5)&\subset &Spin(6) &\subset &Spin(7) & \subset & SO(8)=Spin(8)/Z_2 \\
 \downarrow & & \downarrow & & \downarrow & & \downarrow \\
S^7 & = & S^7 & = & S^7 & = & S^7 \\

 Sp({\bf H}) & & U({\bf C})& & Oct({\bf O}) & & O({\bf R})

\end{array}
\end{equation}

\section{Structure of Special Holonomy manifolds}
 
 We return to a question mentioned above. In the old Kaluza-Klein
 approach, where $M_D \rightarrow M_4$, one gets gauge forces in the lower space from the {\em isometries}
 of the compactifying manifold; for example, the $U(1)$ for the electromagnetic field in the original 
$5 \rightarrow 4$ reduction of Kaluza (1919). But now, where extra dimensions are there to stay, 
the argument does not work anymore! Special holonomy manifolds have, generically, no isometries; 
so if we are to rely on simple gravity in higher spaces, how do we get gauge 
forces in our mundane $4D$ space?

	The answer is spectacular, and I do not think it has been assimilated wholly by the scientific community: 
the special holonomy manifolds have a rich homology, and the non-trivial cycles (that is, the 
uncontractible spheres) can act as sources for gauge fields, following the pattern of singularities,
 $A-D-E$ classification, and Mac-Kay correspondence! \cite{Arnold}, \cite{He}. In a way, this is the
 generalization of the fact that the open string sustains gauge groups in its boundary, the singular points.
 Or, that compactification of M-theory in a segment necessitates two $E_8$ groups in the border, 
the \u{H}orawa-Witten mechanism.

 We do not want to pursue a line of research which seems to be incomplete as yet. However, we cannot
 resist to consider the case of the perhaps simplest special (special but not exceptional) holonomy, 
the case of $K3$ (see an early example in \cite {Duff}), which appears when the $11 \rightarrow 7$ or
 $10 \rightarrow 6$ descents; it is representative of the many sophisticated constructions of 
Joyce and others. So let us construct $K3$ \cite{Aspinwall}

	1) Start with $R^4$, divide by a lattice $L$ to generate a $4$-Torus

\begin{equation}
T^4=R^4/L=R^4/(Z+Z+Z+Z)
\end{equation}

2) Now apply a {\em discrete group $ \Gamma $} with a {\em non-free} action, for example the ``parity"
operation generating $Z_2$:

\begin{equation}
\theta_i \rightarrow -\theta_i 
\end{equation}
for the four angles labelling $T^4$. Call $X=T^4/\Gamma$. The space X is an {\em orbifold}, that is,
a manifold with some special (singular) points, those fixed by $\Gamma$ (here there are $2^4 = 16$ points).
 
3) The $4$-Torus is a complex surface, the X space is
also a complex surface, and as a complex manifold, there is a perfectly standard procedure (starting in 
the 19th century by italian mathematicians!) to {\em remove} (blow-up) the {\em singularities}, trading them,
 in our case, by $2$-spheres (complex projective lines, really). The resulting true, {\it bona fide} 
smooth manifold is called $K3$ in the literature (for Kummer, K\"{a}hler and Kodaira, {\em amen} for 
coincidence with the Himalaya peaks; the godfather seems to be A. Weil \cite{Aspinwall}). 

	The reader can think of converting the cone $x^2 + y^2 = z^2$ on the one-sheeted hyperboloid $x^2 + y^2
-z^2 =1$ as a simple blow-up of the conic singularity, trading it by a circle.

	It is useful to pursue the change in topology: the Betti numbers are

$(1,0, 0, 0, 0)$ for $R^4$, $(1,4,6,4,1)$ for $T^4$, $(1,0,6,0,1)$ for $X$ and $(1,0,22,0,1)$ for $K3$.

	Notice the change after the blow-up: each of the $16$ singular points {\em fattens} to become a ``hollow"
$2$-sphere, hence $b_2$ increases from $6$ to $22$; notice also the increase in "curvature" from $T^4$, 
which is flat: already in $X$ there is ``point" curvature.

	From the point of view of \underline{string theory}, the point of introducing $K3$ is that string theory
$IIA$ or $IIB$ dualizes with the heterotic string in this curious way, in six dimensions \cite{Witten}:

\begin{equation}
II/K3 \approx  \mbox{Het}/T^4
\end{equation}

	In other words, $K3$ ``generates" whatever remains in $6D$ of the $496$-dim gauge group extant in $10D$!
We do not enter into details, as they are well known, albeit not well understood.

\section{Acknowledgements}
This is a slightly worked out edition of the talk I gave in Salamanca.
It is meant to recall the past happy times forty years ago when Adolfo and myself were together in
Barcelona learning the rudiments of cohomology (mainly from Prof. Juan Sancho Guimera), with 
applications to projective representations and group extensions.

\end{document}